\documentclass[aps,prc,twocolumn,showpacs,floatfix]{revtex4-1}

\usepackage{graphicx}% Include figure files
\usepackage{dcolumn}% Align table columns on decimal point
\usepackage{bm}% bold math
\usepackage{ulem}
\usepackage{color}

\newcommand{\al}{$\alpha$}
\newcommand{\g}{$\gamma$}

\newcommand{\ran}{($\alpha$,n)}

\newcommand{\rgn}{($\gamma$,n)}
\newcommand{\rng}{(n,$\gamma$)}

\newcommand{\rdp}{(d,p)}

\newcommand{\Nsv}{$N_A$$\left< \sigma v \right>$}
\newcommand{\sigDC}{$\sigma^{DC}$}
\newcommand{\beviii}{$^{8}$Be}
\newcommand{\beix}{$^{9}$Be}
\newcommand{\benull}{$^{10}$Be}
\newcommand{\cii}{$^{12}$C}

\begin{document}

\title{Direct capture cross section of $^{9}$Be(n,$\gamma$)$^{10}$Be
}

\author{Peter Mohr} \email[Electronic Address: ]{mohr@atomki.mta.hu}
\affiliation{Diakonie-Klinikum, Schw\"{a}bisch Hall D-74523, Germany}
\affiliation{Institute for Nuclear Research (Atomki), Debrecen H-4001, Hungary}
\date{\today}

\begin{abstract}
The cross section of the $^{9}$Be(n,$\gamma$)$^{10}$Be reaction was calculated
in the direct capture model. All parameters of the calculations were adjusted
to properties of the $^{9}$Be + n system at thermal energies. The calculated
cross section at thermonuclear energies shows the expected $1/v$ behavior of
$s$-wave capture at low energies, but increases towards higher energies as
typical $p$-wave capture. Excellent agreement between new experimental data in
the astrophysically relevant energy region and the present calculation is
found.
\end{abstract}

\maketitle

\section{Introduction}
\label{Sec:intro}
In a recent study the \beix \rng \benull\ reaction was investigated at thermal
and stellar energies \cite{Wallner_PRC2019_be9ng}. The main aim of that study
was the measurement of the cross section at energies in the keV region which
is essential to determine the astrophysical reaction rate at the high
temperatures which can be found during core-collapse supernova
explosions. Here the \beix \rng \benull\ reaction may play an important role
in the so-called \al -process under neutron-rich conditions
\cite{Wallner_PRC2019_be9ng}.

In general, the formation of \cii\ from nucleons and \al\ particles is
hindered by the gaps of stable nuclei at masses $A = 5$ and $A = 8$ which has
to be bypassed by three-particle reactions. Depending on the \al\ and neutron
densities in the astrophysical environment, the triple-alpha (\al \al \al )
process may be supplemented by the (\al \al n) or (\al nn) reactions which
both proceed via \beix , either directly produced in (\al \al n) or indirectly
in (\al nn) and subsequent $^{6}$He\ran \beix . Then \cii\ can be formed from
the \beix \ran \cii\ reaction; however, \beix\ can also be detracted from the
\cii\ formation by either the \beix \rng \benull\ or \beix \rgn
\beviii\ reactions (the latter becoming only relevant at high
temperatures). The neutron-rich bypasses to the triple-alpha process occur in
the \al -process in core-collapse supernovae. The onset of the \al -process is
discussed in detail in \cite{Woosley_APJ1992_alpha}, and further information
on the relevance of the different three-body processes is given in
\cite{Goerres_PRC1995_he4nn,Bartlett_PRC2006_he4nn}.

Experimental data for the \beix \rng \benull\ reaction in the keV region are
very sparse. The resonance properties of the lowest resonance in \beix \rng
\benull\ have been studied by Kitazawa {\it et
  al.}\ \cite{Kitazawa_NPA1994_be9ng}, and three data points with relatively
large error bars are provided by Shibata in an unpublished thesis made in the
same group \cite{Shibata_thesis1992}. This gap is filled now by the new
experimental data of Wallner {\it et al.}\ \cite{Wallner_PRC2019_be9ng}.

A very brief theoretical analysis of the new experimental data in the direct
capture model is also given in \cite{Wallner_PRC2019_be9ng}, and it is
concluded that the $p$-wave contribution had to be scaled down by about 30\%
to fit the new experimental data. It is the scope of the present study to
provide a more detailed analysis of the direct capture process in the \beix
\rng \benull\ reaction. It will be shown that the new data in the keV region
can be well described if the parameters of the calculation are carefully
chosen to reproduce the well-known properties of \beix\ + n at thermal
energies (i.e., without any additional adjustment of parameters to the new
data in the keV region). Furthermore, the contribution of low-lying resonances
is re-analyzed, leading to a slightly different reaction rate at very high
temperatures. Obviously, there is no major change in the astrophysical
reaction rate at lower temperatures because finally the calculated $p$-wave
contributions in \cite{Wallner_PRC2019_be9ng} (adjusted to fit the new data in
the keV region) and in this study (which fit the keV data without adjustment)
are practically identical.

\section{The direct capture model}
\label{sec:DC}
\subsection{Basic considerations}
\label{sec:basic}
As long as the level density in the compound nucleus (\benull\ in the present
case) is low, resonances play only a minor role, and the capture cross section
is dominated by the direct capture (DC) process. Often this is the case
for light nuclei, but DC may also be dominant for neutron-rich nuclei, in
particular with closed neutron shells, where the low $Q$-value of neutron
capture corresponds to relatively small excitation energies and thus low level
densities in the compound nucleus. As a nice example, DC was experimentally
confirmed for the $^{48}$Ca\rng $^{49}$Ca reaction \cite{Beer_PRC1996_ca48ng},
and it was possible to describe the cross section in the keV region after
adjustment of the parameters to thermal properties of the $^{48}$Ca + n system.

The full DC formalism is given by Kim {\it et al.}\ \cite{Kim_PRC1987_DC} and
also listed in \cite{Beer_PRC1996_ca48ng,Mohr_PRC1998_mg26ng}. Basic
considerations on DC have already been provided by Lane and Lynn more than 50
years ago \cite{Lane_NP1960_DC1,Lane_NP1960_DC2}. The chosen
model in \cite{Wallner_PRC2019_be9ng} is based on
\cite{Mengoni_PRC1995_p-capture} which contains the same underlying physics
with a focus on direct $p$-wave capture. Here I briefly repeat only the
essential features of the DC model; for details, see
\cite{Beer_PRC1996_ca48ng,Mohr_PRC1998_mg26ng,Mengoni_PRC1995_p-capture,Kim_PRC1987_DC}.

The DC cross sections \sigDC\ scale with the square of the overlap integrals
$\cal{I}$
\begin{equation}
{\cal{I}} = \int dr \, u(r) \, {\cal{O}}^{E1/M1} \, \chi(r)
\label{eq:overlap}
\end{equation}
where ${\cal{O}}^{E1/M1}$ is the electric or magnetic dipole operator; E2
transitions are much weaker than E1 transitions for the light $N \ne Z$
nucleus \benull\ and can be neglected for the DC calculations. The $u(r)$ and
$\chi(r)$ are the bound state wave function and scattering state wave
function. These wave functions are calculated from the two-body Schr\"odinger
equation using a nuclear potential without imaginary part because the damping
of the wave function in the entrance channel by the small DC cross sections is
typically very small \cite{Krausmann_PRC1996}. Finally, the DC cross section
has to be normalized with the spectroscopic factor $C^2 S$ to obtain the
capture cross section $\sigma_{\gamma,f}$ to a final state $f$:
\begin{equation}
\sigma_{\gamma,f} = (C^2 S)_f \, \sigma^{DC}_f \quad .
\label{eq:SF}
\end{equation}
The total capture cross section $\sigma_{\gamma}$ is obtained by the sum over
all final states $f$:
\begin{equation}
\sigma_{\gamma} = \sum_f \sigma_{\gamma,f} \quad .
\label{eq:sum}
\end{equation}

An essential ingredient for the DC model is the nuclear potential $V(r)$ for
the calculation of the wave functions $u(r)$ and $\chi(r)$. In the present
work, a folding potential was used:
\begin{equation}
  V(r) = \lambda \, V_F(r)
  \label{eq:fold}
\end{equation}
with the strength parameter $\lambda$ of the order of unity. For details of
the folding potential, see \cite{Beer_PRC1996_ca48ng,Mohr_PRC1998_mg26ng}. The
advantage of the folding potential is that only one parameter, namely the
strength $\lambda$, has to be adjusted which reduces the available parameter
space significantly (compared to the widely used Woods-Saxon potentials with
three parameters).

\subsection{Adjustment of the potential}
\label{sec:pot_adjust}
For the calculation of bound state wave functions $u(r)$, the potential
strength is adjusted to the binding energy of the respective state to ensure
the correct asymptotic shape of $u(r)$. Thus, the only parameter $\lambda$ of
the potential is fixed for each final state $f$, and all wave functions $u(r)$
can be calculated without further adjustment of parameters (see Table
\ref{tab:bound}).
\begin{table*}[htb]
  \caption{
    \label{tab:bound}
Bound state properties of states in \benull\ below the neutron
threshold. Energies $E_x$ and spins and parities $J^\pi$ are taken from
\cite{Tilley_NPA2004_A8-10}. The effective spectroscopic factors $C^2
S_{\rm{eff}}$ of this study are defined in the text. Spectroscopic factors
from transfer reactions are taken from the ENSDF database \cite{ENSDF} and the
compilation of Tilley {\it et al.}\ \cite{Tilley_NPA2004_A8-10}.
}
\begin{center}
\begin{tabular}{rrccrrrrrr}
\hline
\multicolumn{1}{c}{$E_x$}
& \multicolumn{1}{c}{$E_B$} 
& \multicolumn{1}{c}{$J^\pi$} 
& \multicolumn{1}{c}{$L$} 
& \multicolumn{1}{c}{$\lambda$} 
& \multicolumn{1}{c}{$C^2 S_{\rm{eff}}$}
& \multicolumn{1}{c}{$C^2 S$\rdp } 
& \multicolumn{1}{c}{$C^2 S$(\al ,$^3$He)} 
& \multicolumn{1}{c}{$C^2 S$($^7$Li,$^6$Li)}
& \multicolumn{1}{c}{$C^2 S$($^8$Li,$^7$Li)} \\
\multicolumn{1}{c}{(MeV)}
& \multicolumn{1}{c}{(MeV)} 
& \multicolumn{1}{c}{--} 
& \multicolumn{1}{c}{--} 
& \multicolumn{1}{c}{--} 
& \multicolumn{1}{c}{--} 
& \multicolumn{1}{c}{--} 
& \multicolumn{1}{c}{--} 
& \multicolumn{1}{c}{--} 
& \multicolumn{1}{c}{--} \\
\hline
$0.0$     & $-6.8123$ & $0^+$ & $1$ & 1.1311 & 1.794
& $\approx 1.06$ & 1.58~ & 2.07 & 4.0 \\
$3.3680$  & $-3.4443$ & $2^+$ & $1$ & 0.9601 & 2.963
& 0.17           & 0.38~ & 0.42 & 0.2 \\
$5.9584$  & $-0.8539$ & $2^+$ & $1$ & 0.7986 & 1.357
& 0.54           & $< 0.73$\footnote{unresolved} & -- & -- \\
$5.9599$  & $-0.8524$ & $1^-$ & $0$ & 1.3492 & 0.523
& --             & $< 0.14$\footnotemark[1] & -- & -- \\
$6.1793$  & $-0.6330$ & $0^+$ & $1$ & 0.7818 & 0.048
& --             & -- & -- & -- \\
$6.2633$  & $-0.5490$ & $2^-$ & $0$ & 1.3084 & 0.467
& --             & 0.08~ & -- & -- \\
\hline
\end{tabular}
\end{center}
\end{table*}

The scattering wave function $\chi(r)$ for the $s$-wave with angular momentum
$L = 0$ has to reproduce the thermal scattering length. From the bound
coherent and incoherent scattering lengths $b_c = 7.79 \pm 0.01$ fm and $b_i =
0.12 \pm 0.03$ fm \cite{Sears_NeutronNews1992} it turns out that the free
scattering lengths $a_{+}$ and $a_{-}$ for $J^\pi_{+} = 2^-$ and $J^\pi_{-} =
1^-$ are almost identical, and thus for simplicity a weighted average
$\lambda = 1.4159$ was used for all scattering $s$-waves instead of
$\lambda_{+} = 1.4097$ and $\lambda_{-} = 1.4263$. Note that the above
$J^\pi_{+}$ and $J^\pi_{-}$ result from the coupling of the neutron spin
$I^\pi_n = 1/2^+$, the spin of the \beix\ ground state $I^\pi_T = 3/2^-$, and
angular momentum $L = 0$. The very minor variations of $\lambda$ within about
1\% do practically not affect the calculated DC cross sections.

The adjustment of the potential strength for the scattering $p$-wave is more
complicated because the thermal scattering lengths are related to $s$-wave
scattering only. As an alternative, the same procedure as for the bound states
was applied. Parameters $\lambda$ were determined by adjustment to all bound
($E < 0$) and quasi-bound ($E > 0$) states in \benull\ where $L = 1$ transfer
was clearly assigned in the \beix \rdp \benull\ or \beix (\al
,$^{3}$He)\benull\ reactions \cite{Tilley_NPA2004_A8-10}. From the average of
all $L = 1$ states one finds a significantly lower $\lambda = 0.8856$ for the
$p$-wave, compared to $\lambda =1.4159$ for the $s$-wave. Similar to the
$s$-wave, the same value for $\lambda$ was used for both channel spins
$J^\pi_{+}$ and $J^\pi_{-}$.

\subsection{Adjustment of spectroscopic factors}
\label{sec:SF_adjust}
Spectroscopic factors are required for neutron transfer to the $0p_{3/2}$,
$0p_{1/2}$, and $1s_{1/2}$ shells. As the potential $V(r)$ is well-constrained
for the incoming $s$-wave at thermal energies, spectroscopic factors $C^2 S$
can be derived from the thermal neutron capture cross section of \beix\ using
Eq.~(\ref{eq:SF}).

The thermal neutron capture cross section has been
determined in several experiments, and the results are in excellent
agreement. I adopt $\sigma_{\gamma} = 8.299 \pm 0.119$ mb which results from
the weighted average of
$8.27 \pm 0.13$ mb \cite{Firestone_PRC2016_ng},
$8.49 \pm 0.34$ mb \cite{Conneely_NIMA1986_be9ng}, and
$8.31 \pm 0.52$ mb from the new experiment \cite{Wallner_PRC2019_be9ng}. The
branching ratios to the individual final states in \benull\ are also taken
from the recent experiment by Firestone and Revay \cite{Firestone_PRC2016_ng}.

For the bound states with $J^\pi = 2^+$, contributions of the transfers to the
$0p_{3/2}$ and $0p_{1/2}$ shells have to be added. However, this can be
simplified because the $s$-wave capture scales approximately with $1/v$ for
any combination of $0p_{3/2}$ and $0p_{1/2}$ transfer. As long as a proper
adjustment to the capture cross section is made at thermal energies, the
$s$-wave capture in the keV region must also be reproduced. Therefore, an
effective spectroscopic factor $C^2 S_{\rm{eff}}$ is listed in Table
\ref{tab:bound} which takes into account only the transfer to the $0p_{3/2}$
shell; contributions of the $0p_{1/2}$ transfer are neglected.

The adjustment of the effective spectroscopic factors to the thermal capture
cross section is fortunately possible also for the $2^-$ state at $E_x =
6.263$ MeV because a weak M1 transition to this state was detected in
\cite{Firestone_PRC2016_ng}. Only for the $1^-$ state at $E_x = 5.960$ MeV an
adjustment of $C^2 S$ from the thermal capture cross section is not possible
because no primary \g -ray could be detected. Consequently, $C^2 S$ for this
state had to be fixed in a different way. For that purpose a procedure was
used which relates the thermal scattering lengths to the spectroscopic factors
of subthreshold $s$-wave states \cite{Mohr_PRC1997_SF}. As the adjusted $C^2
S$ for the neighboring $2^-$ state from Eq.~(\ref{eq:SF}) is about 35\% lower
than $C^2 S$ from the procedure of \cite{Mohr_PRC1997_SF}, the same reduction
factor was applied for the unknown $C^2 S$ for the $1^-$ state, leading to
$C^2 S = 0.523$ (see Table \ref{tab:bound}). This value is roughly consistent
with $C^2 S \approx 0.4$ which can be derived with huge uncertainties from a
weak secondary \g -ray in thermal neutron capture after correction for feeding
\cite{Firestone_PRC2016_ng}.

A comparison of the effective spectroscopic factors $C^2 S_{\rm{eff}}$ in
Table \ref{tab:bound} to spectroscopic factors from transfer reactions like
\beix \rdp \benull\ is not straightforward. First, the effective spectroscopic
factors of this study are calculated for the transfer to the $0p_{3/2}$ shell
only which simplifies the present calculations (see discussion above), but
complicates the comparison to data from transfer reactions. Second,
spectroscopic factors from transfer depend on the chosen parameters of the
underlying calculations of the reaction cross sections
\cite{Mukhamedzhanov_PRC2008_d-p}, which are typically based on the distorted
wave Born approximation (DWBA). This is reflected by wide variations of $C^2
S$ from \rdp , (\al ,$^3$He), and ($^7$Li,$^6$Li). In some cases there is even
disagreement on the transferred angular momentum $L$. The generally poor
agreement of the $C^2 S$ from different transfer reactions is explicitly
stated in the compilation of Tilley {\it et al.}\ \cite{Tilley_NPA2004_A8-10}.
Third, the two levels around $E_x = 5.96$ MeV in \benull\ cannot be resolved
easily in transfer experiments. Therefore, I restrict myself here to list the
adopted spectroscopic factors from different transfer reactions in Table
\ref{tab:bound} (as compiled in the ENSDF database \cite{ENSDF} or given in
Tilley {\it{et al.}} \cite{Tilley_NPA2004_A8-10}). The only noticeable
peculiarity is the deviation for the first excited $2^+$ state in
\benull\ between the huge $C^2 S_{\rm{eff}} \approx 3.0$ from the thermal
\rng\ cross section and $C^2 S \approx 0.2 - 0.4$ from different transfer
reactions. The thermal branching to the $2^+$ state at $E_x = 3.368$ MeV is
moderate with about 11\%, but well-defined \cite{Firestone_PRC2016_ng}, and
thus $C^2 S_{\rm{eff}}$ is well-constrained in the present approach. A more
detailed discussion of spectroscopic factors is omitted because of the
significant uncertainties of the $C^2 S$ from the different transfer
reactions.

\section{Results and discussion}
\label{sec:res}
\begin{figure}[htb]
\includegraphics[width=\columnwidth,clip=]{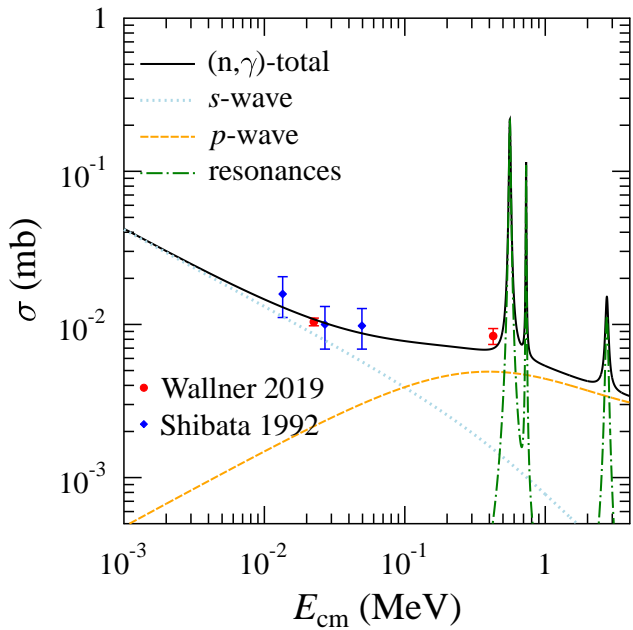}
\caption{
\label{fig:sigma}
(Color online)
Cross section of the \beix \rng \benull\ reaction. The calculated total cross
section (full black line) is composed of $s$-wave (lightblue dotted) and
$p$-wave (orange dashed) DC plus resonances (green dash-dotted). Excellent
agreement with the experimental data
\cite{Wallner_PRC2019_be9ng,Shibata_thesis1992} is found. Further discussion
see text.
}
\end{figure}

After the adjustment of the potential in Sec.~\ref{sec:pot_adjust} and of the
spectroscopic factors in Sec.~\ref{sec:SF_adjust}, all parameters for the DC
calculations are now completely fixed. The DC cross sections for $s$-wave and
$p$-wave capture can now be calculated without any further adjustment of
parameters. The results are shown in Fig.~\ref{fig:sigma}. As usual, $s$-wave
capture decreases with energy by roughly $1/\sqrt{E}$, whereas $p$-wave
capture increases with $\sqrt{E}$. A transition from the $1/\sqrt{E}$ to the
$\sqrt{E}$ behavior is found at several tens of keV. This is a typical result
for light nuclei at the upper end of the $p$-shell like $^{12}$C
\cite{Ohsaki_APJ1994_c12ng} and in the $sd$-shell (e.g., $^{16}$O
\cite{Igashira_APJ1995_o16ng,Mohr_APJ2016_o16ng} and $^{26}$Mg
\cite{Mohr_PRC1998_mg26ng}).

\begin{figure}[htb]
\includegraphics[width=\columnwidth,clip=]{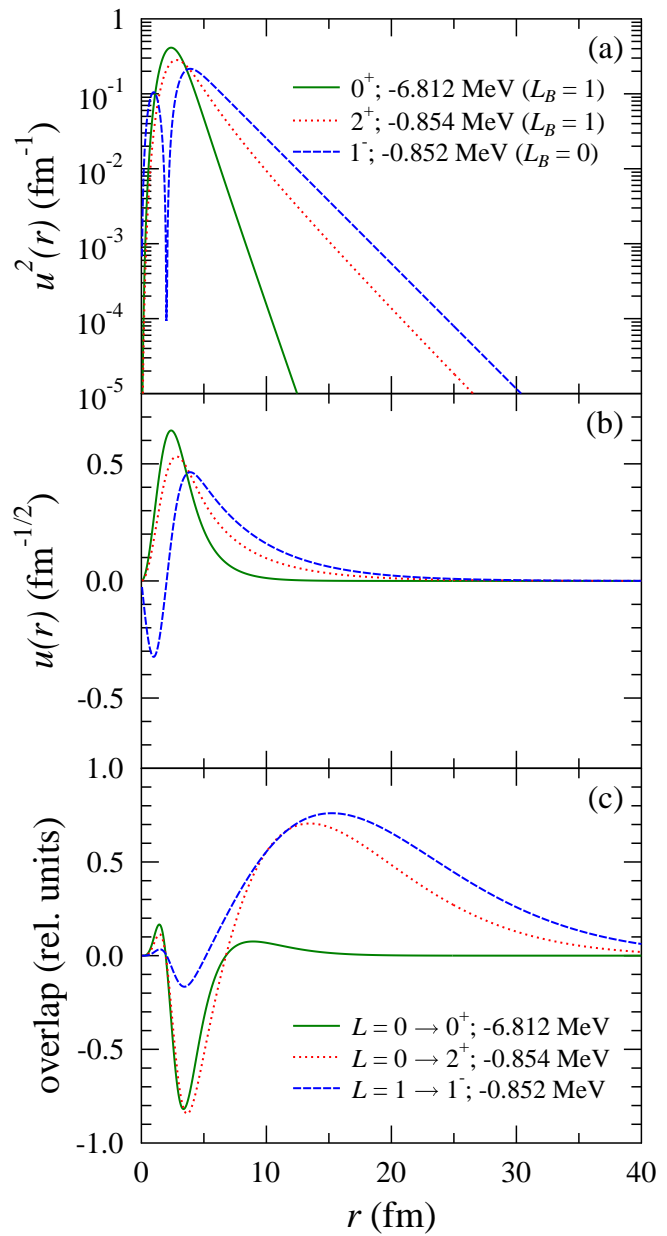}
\caption{
\label{fig:wave}
(Color online) Wave functions $u^2(r)$ (upper part a, logarithmic scale) and
$u(r)$ (middle part b, linear scale) of the $0^+$ ground state and the $2^+$
and $1^-$ dublett of states at $E_x = 5.96$ MeV (see also Table
\ref{tab:bound}). The resulting integrand of the overlap integral in
Eq.~(\ref{eq:overlap}) is also shown for $E = 100$ keV (lower part c). Further
discussion see text.
}
\end{figure}

Important ingredients of the DC calculations like wave functions and overlaps
are further illustrated in Fig.~\ref{fig:wave}. Both bound state wave
functions $u(r)$ (shown in the upper part a as $u^2(r)$ in logarithmic scale
and in the middle part b as $u(r)$ in linear scale) of the $0^+$ ground state
and the $2^+$ excited state at 5.96 MeV are characterized by $L_B = 1$ and
thus mainly differ in the exterior which is determined by the binding energies
of both states. Contrary, the $1^-$ state at 5.96 MeV has $L_B = 0$ and one
node in the interior. In the exterior, the $2^+$ and $1^-$ wave functions show
the same slope because of the almost identical binding energies.

The resulting integrand of the overlap integral in Eq.~(\ref{eq:overlap}) is
shown in the lower part c of Fig.~\ref{fig:wave} for a chosen energy $E = 100$
keV. Obviously, the main contributions for the capture to the ground state
come from the nuclear interior and surface at relatively small radii ($r
\lesssim 10$ fm). Because of the smaller binding energies of the $2^+$ and
$1^-$ final states, the main contributions for the transitions to these states
appear in the nuclear exterior for radii 10 fm $\lesssim r \lesssim 40$
fm. Nevertheless, for all transitions noticeable cancellation effects are
found between the positive and the negative areas of the integrands in
Fig.~\ref{fig:wave} (part c). A similar observation has already been made in
an earlier study of direct neutron capture at thermal energies
\cite{Lynn_PRC1987_ng} which is based on the model described in
\cite{Raman_PRC1985_s_ng}.

The DC calculation of $s$-wave and $p$-wave capture is complemented by the
contributions of the four lowest known resonances which correspond to the
states in \benull\ at $E_x = 7.371$ MeV ($J^\pi = 3^-$), 7.542 MeV ($2^+$),
9.27 MeV ($4^-$), and 9.56 MeV ($2^+$). The properties of the resonances are
listed in Table \ref{tab:res}.
\begin{table}[htb]
  \caption{
    \label{tab:res}
    Properties of low-energy resonances in the \beix \rng \benull\ reaction
    (taken from \cite{Tilley_NPA2004_A8-10} except $\Gamma_\gamma$). The
    estimates of the radiative widths $\Gamma_\gamma$ are explained in the
    text. All energies and widths are given in the c.m.\ system.
}
\begin{center}
\begin{tabular}{rrcccc}
\hline
\multicolumn{1}{c}{$E_x$}
& \multicolumn{1}{c}{$E_R$} 
& \multicolumn{1}{c}{$J^\pi$} 
& \multicolumn{1}{c}{$L$}
& \multicolumn{1}{c}{$\Gamma$} 
& \multicolumn{1}{c}{$\Gamma_\gamma$} \\
\multicolumn{1}{c}{(MeV)}
& \multicolumn{1}{c}{(MeV)} 
& \multicolumn{1}{c}{--} 
& \multicolumn{1}{c}{--} 
& \multicolumn{1}{c}{(keV)} 
& \multicolumn{1}{c}{(eV)} \\
\hline
$7.371$  & $0.559$ & $3^-$ & $2$ & 15.7 & 0.73
\footnote{experimental value taken from \cite{Kitazawa_NPA1994_be9ng}}\\
$7.542$  & $0.730$ & $2^+$ & $1$ & 6.3 & 0.28
\footnote{estimated from average radiation widths} \\
$9.270$  & $2.458$ & $(4^-)$ & $2$   & 150 & 1.3 meV
\footnotemark[2] \\
$9.56$   & $2.748$ & $2^+$   & $1$  & 141 & 2.43
\footnotemark[2] \\
\hline
\end{tabular}
\end{center}
\end{table}

For the calculation of the resonance cross sections the approximation $\Gamma
\approx \Gamma_n$ was used because it is known that $\Gamma_\alpha / \Gamma_n
\ll 0.1$ for these states \cite{Tilley_NPA2004_A8-10}. The radiation width of
the lowest resonance was determined experimentally as $\Gamma_\gamma = 0.73$
eV \cite{Kitazawa_NPA1994_be9ng}. This $3^-$ resonance decays by E1
transitions to the first excited state in \benull\ ($\Gamma_\gamma = 0.62 \pm
0.06$ eV which corresponds to a noticeable strength of 31 mW.u.) and to the
second excited state ($\Gamma_\gamma = 0.11 \pm 0.08$ eV, corresponding to 124
mW.u.).

If one assumes the same average Weisskopf units for the E1 transitions in the
decay of the next resonance with $J^\pi = 2^+$ at $E_x = 7.542$ MeV, one ends
up with a smaller radiation width of $\Gamma_\gamma \approx 0.14$ eV because
E1 transitions can only lead to odd-parity states around $E_x \approx 6$ MeV
and thus correspond to relatively low transition energies. Because of the high
transition energy of the E2 transition to the ground state, almost the same
radiation width for the E2 transition can be estimated using a typical
strength of about 5 W.u.\ for E2 transitions in this mass region
\cite{Endt_ADNDT1993_A5-44}. This leads to an overall radiation width of
$\Gamma_\gamma = 0.28$ eV which is significantly lower than assumed by Wallner
{\it et al.}\ who use the same $\Gamma_\gamma = 0.73$ eV as for the $3^-$
resonance.

Assuming the same strengths of 75 mW.u.\ for E1 and 5 W.u.\ for E2
transitions, the $(4^-)$ resonance has only a tiny radiation width of
$\Gamma_\gamma \approx 1.3$ meV which results from the E2 transition to the
$2^-$ state at $E_x = 6.263$ MeV. Additional \g -transitions may occur to the
levels in \benull\ above the neutron threshold with larger strength (e.g., for
the M1 transition to the $3^-$ state at 7.371 MeV); however, the final state
of this transition decays preferentially by neutron emission and thus does not
contribute to \benull\ production.

A large radiation width is found for the $2^+$ state at 9.56 MeV because of
strong E2 transitions to low-lying $0^+$ and $2^+$ states in \benull :
$\Gamma_\gamma \approx 2.43$ eV. However, this resonance is located at almost
3 MeV and thus contributes to the astrophysical reaction rate only at very
high temperatures.

Interference effects between the resonances are not taken into account in the
present study because no experimental information is available. However, it
can be estimated that interference effects will be minor because the
dominating $3^-$ $d$-wave resonance does not interfere with the dominating
$p$-wave DC contributions.

For completeness it has to be noted that the two $L = 2$ resonances contain a
significant amount of the total $L = 2$ strength. As these resonances are
taken into account explicitly, an additional calculation of the $d$-wave
contribution of the DC cross section would double-count the $L = 2$ strength,
and thus the $d$-wave contribution of the DC cross section is intentionally
omitted. The folding potential for the $s$-waves contains two $L = 0$ bound
states close below the neutron threshold (see Table \ref{tab:bound}). Assuming
the same potential for the $d$-wave automatically leads to the appearance of
$d$-wave resonances at low energies which are the theoretical counterparts of
the experimentally observed $d$-wave resonances (see Table \ref{tab:res}).

Overall, the agreement between the calculated total cross section and the new
experimental data \cite{Wallner_PRC2019_be9ng} is very good with a small
$\chi^2 \approx 1.25$ per point. The dominating contribution to $\chi^2$ comes
from the upper data point at $E_{n,{\rm{lab}}} = 473 \pm 53$ keV where an
average cross section of $\sigma_{\rm{exp}}^{\rm{av}} = 8.4 \pm 1.0$ $\mu$b is
reported in \cite{Wallner_PRC2019_be9ng}. The calculated cross section at
exactly 473 keV is $\sigma_{\rm{calc}} = 6.97$ $\mu$b. Averaging the
calculated cross section over the experimental energy distribution of the
neutrons (see Fig.~3 of \cite{Wallner_PRC2019_be9ng}) leads to
$\sigma_{\rm{calc}}^{\rm{av}} = 7.18$ $\mu$b which deviates only by
1.2$\sigma$ from the experimental $\sigma_{\rm{exp}}^{\rm{av}}$.  The increase
from 6.97 $\mu$b to 7.18 $\mu$b results from the higher calculated cross
sections at the upper end of the experimental neutron energy interval. As a
consequence, $\chi^2$ per point approaches 1.0 in this case.  Including the
Shibata points \cite{Shibata_thesis1992} reduces the deviations further to
$\chi^2 \approx 0.6$ per 
point. It has to be repeated that the present calculation has been made
completely independent, without any adjustment to the new experimental data
points in the keV region.

\section{Astrophysical reaction rate}
\label{sec:rate}
\begin{figure}[ht]
\includegraphics[width=\columnwidth,clip=]{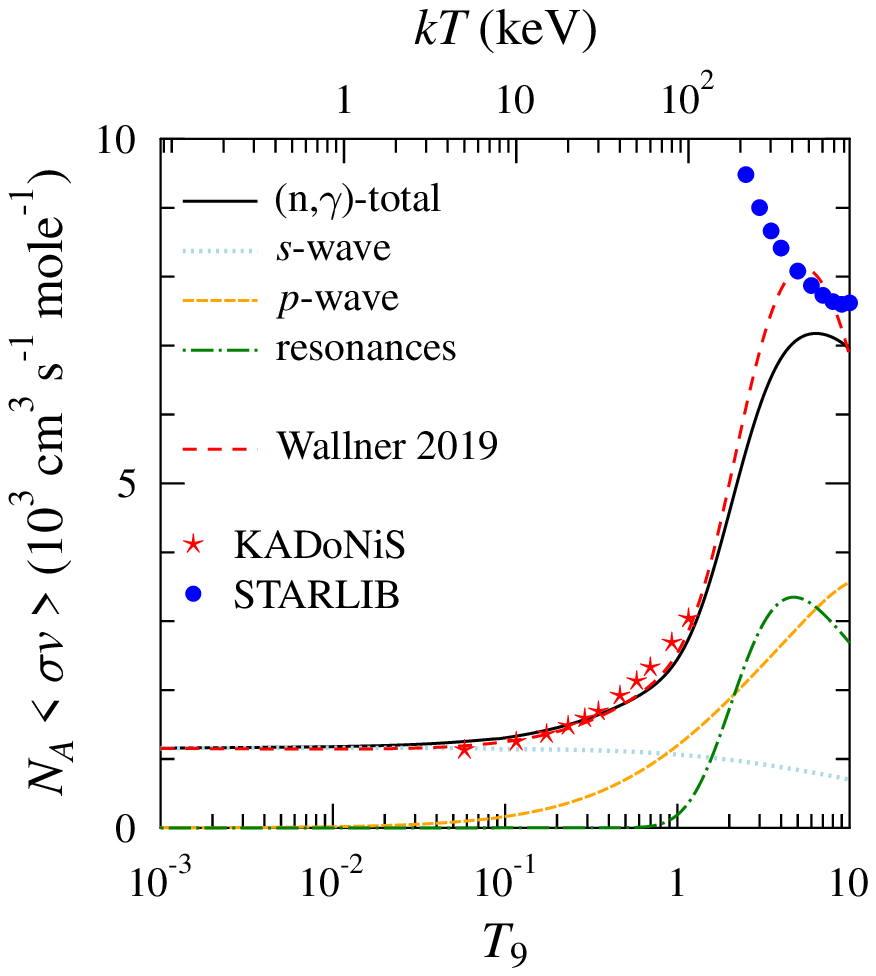}
\caption{
\label{fig:rate}
(Color online)
Astrophysical reaction rate \Nsv\ for the production of \benull\ in the \beix
\rng \benull\ reaction.
}
\end{figure}

The astrophysical reaction rate \Nsv\ was calculated by numerical integration
of the cross sections in Sec.~\ref{sec:res}. A narrow energy grid from 1 to
4000 keV was used to cover the the full temperature range up to $T_9 =
10$. Because of the relatively high first excited state in \beix , no stellar
enhancement factor was used (as also suggested in the KADoNiS database
\cite{KADONIS}). The result is shown in Fig.~\ref{fig:rate}.

At low temperatures below a few keV this energy grid is not
sufficient. Therefore, the calculation of the cross section was repeated in 10
eV steps from 10 eV to 50 keV. With these settings a constant rate for the
$s$-wave capture was found down to the lowest temperatures in
Fig.~\ref{fig:rate} which confirms that the numerical treatment is stable.

The $s$-wave capture dominates the low-temperature region below $T_9 \approx
0.1$ whereas at higher temperatures around $T_9 \approx 1$ $p$-wave capture
becomes the major contributor. At even higher temperatures the resonance
contributions become comparable to $p$-wave capture which result mainly from
the lowest $3^-$ resonance at 559 keV.

As expected, the present rate is in very good agreement with the rate by
Wallner {\it et al.}\ \cite{Wallner_PRC2019_be9ng} because their DC
calculation was adjusted to their new experimental data (whereas the present
calculation reproduces the new experimental data without adjustment). The only
significant difference appears at relatively high temperatures around $T_9
\approx 5$ and results from the lower resonance strength of the lowest $2^+$
resonance in the present study (see Table \ref{tab:res} and discussion in
Sec.~\ref{sec:res}). At the highest temperature $T_9 = 10$ in
Fig.~\ref{fig:rate} the present rate becomes similar to the Wallner rate again
because the lower strength of the $2^+$ resonance is compensated by the
additional resonances at higher energies which were not taken into account in \cite{Wallner_PRC2019_be9ng}. 

Fig.~\ref{fig:rate} also includes the recommended rate of the KADoNiS database
\cite{KADONIS} (version 1.0) which was derived from preliminary data of
Wallner {\it et al.}\ and thus can be recommended for astrophysical
calculations. The REACLIB database \cite{Cyburt_APJS2010_reaclib} also
recommends to use the KADoNiS rate. However, STARLIB
\cite{Sallaska_APJS2013_starlib} contains a theoretical rate which is based on
the statistical model. This theoretical rate exceeds the recommended rate by
far at low temperatures and shows a completely different temperature
dependence (see Fig.~\ref{fig:rate}). Such a discrepancy is not very
surprising because the statistical model is inappropriate for such light
nuclei. A comparison of the new capture data to different libraries for
neutron cross sections was already given in \cite{Wallner_PRC2019_be9ng} and
is omitted here.

The astrophysical reaction rate \Nsv\ was fitted using the same
parametrization as in Eq.~(7) of \cite{Wallner_PRC2019_be9ng}:
\begin{eqnarray}
\frac{N_A < \sigma v >}{{\rm{cm}}^3 {\rm{s}}^{-1} {\rm{mol}}^{-1}} = 
  & & \, \, a_0 \,
      \Bigl( \, 1.0+ a_1 T_9^{1/2} + a_2 T_9 + a_3 T_9^{3/2}  
\nonumber \\ 
  & & \, \, \, \, \, \, \, \, \, \, \, \, \, \,
      + a_4 T_9^{2} + a_5 T_9^{5/2} \, \Bigr)
\nonumber \\
  & & \, \, \, + a_6 T_9^{-3/2} \exp{(-b_0/T_9)}
\label{eq:rate_fit}
\end{eqnarray}
The $a_i$ and $b_0$ parameters are listed in Table \ref{tab:rate_fit}. The
deviation of the fitted rate is below 1\% over the full temperature
range.
\begin{table*}[htb]
\caption{\label{tab:rate_fit}
Fit parameters $a_i$ and $b_0$ of the recommended reaction rate \Nsv
$_{\rm{rec}}$ from Eq.~(\ref{eq:rate_fit}).
}
\begin{center}
\begin{tabular}{cccccccc}
\hline
\multicolumn{1}{c}{$a_0$}
& \multicolumn{1}{c}{$a_1$} 
& \multicolumn{1}{c}{$a_2$} 
& \multicolumn{1}{c}{$a_3$} 
& \multicolumn{1}{c}{$a_4$} 
& \multicolumn{1}{c}{$a_5$} 
& \multicolumn{1}{c}{$a_6$}
& \multicolumn{1}{c}{$b_0$} \\
\hline
   1164.44
& -0.062597
&  1.7712
&  -0.95285
&  0.22494
& -0.021126
& 127297.1
& 6.64777 \\
\hline
\end{tabular}
\end{center}
\end{table*}

\section{Conclusions}
\label{sec:conc}
The cross section of the \beix \rng \benull\ reaction was calculated in the
direct capture model. All parameters of the calculations could be adjusted to
thermal properties of the \beix\ + n system, and therefore the calculation of
the capture cross sections in the astrophysically relevant keV region is
completely 
free of any adjustments. The calculated cross sections agree very well with
the recently published experimental results by Wallner {\it et
  al.}\ \cite{Wallner_PRC2019_be9ng} and also with earlier unpublished data by
Shibata \cite{Shibata_thesis1992}. The astrophysical reaction rate of the
KADoNiS database is essentially confirmed; it is based on a preliminary
analysis of the Wallner {\it et al.}\ data. REACLIB also suggests to use the
KADoNiS rate. However, the reaction rate of STARLIB should not be used because
it is based on a statistical model calculation which overestimates the
experimental data significantly.

\acknowledgments 
I thank A.\ Wallner for encouraging discussions.
This work was supported by NKFIH (K108459 and K120666).

\bibliography{pm}

\end{document}